\documentclass[prd, twocolumn, nofootinbib, showkeys]{revtex4}
\usepackage{graphicx}
\usepackage{dcolumn}
\usepackage{epsfig} 

\begin{document} 

\title{Slow-Roll Suppression of Adiabatic Instabilities in Coupled Scalar Field-Dark Matter Models}

\author{Pier Stefano Corasaniti} 
\affiliation{
LUTH, Observatoire de Paris, CNRS UMR 8102, Universit\'e Paris
  Diderot, 5 Place Jules Janssen, 92195 Meudon Cedex, France}
\begin{abstract}
We study the evolution of linear density perturbations in the context of
interacting scalar field-dark matter cosmologies, 
where the presence of the coupling acts as a stabilization mechanism 
for the runaway behavior of the scalar self-interaction potential
as in the case of the Chameleon model. We show that in the
``adiabatic'' background regime of the system the rise of 
unstable growing modes of the perturbations is
suppressed by the slow-roll dynamics of the field. Furthermore the
coupled system behaves as an inhomogeneous adiabatic fluid. In contrast
instabilities may develop for large values of 
the coupling constant, or along non-adiabatic solutions, 
characterized by a period of high-frequency dumped
oscillations of the scalar field. In the latter case the 
dynamical instabilities of the field fluctuations, 
which are typical of oscillatory scalar field regimes, 
are amplified and transmitted by the coupling 
to dark matter perturbations. 
\end{abstract} 

\date{\today}

\maketitle 

\section{Introduction}\label{intro}
Cosmology has provided evidence of a dark physics sector 
which is necessary to account for about $95\%$ of the cosmic matter
content \cite{CMB}. 
Despite the success of the $\Lambda$CDM model to fit all cosmological
observations, the existence of the dark energy phenomenon as well as its relation
to the abundance and clustering of matter in the universe still pose
puzzling questions.

Models of interacting dark energy-dark matter have been
proposed to address such problems. In this scenario
dark energy is a fundamental scalar field which directly couples to
matter particles. This allows for a dynamical solution of the so called
``coincidence'' problem, since independently of the initial 
conditions the scalar interaction drives the dark energy-to-matter
ratio toward a constant value (see e.g. \cite{LucaDomenico,Chimento,Pietroni,JMAndre1}).
These models are inspired by string and supergravity theories, where the compactification 
of extra-dimensions in the low-energy gives rise to massless scalars 
coupled to matter fields with gravitational strength. 
Therefore a distinct feature of this scenario is that
matter particles experience a long-range scalar force and acquire 
a time dependent mass which cause violations of the Equivalence Principle (EP). 
The tight bounds imposed by EP tests are usually avoided as consequence
of other possible mechanisms. As an example Damour and Polyakov have
shown that in String Theory the
couplings between the dilaton and different matter fields can be
dynamically suppressed \cite{Damour}. 
An interesting possibility has been proposed in the
``Chameleon'' model \cite{Justin}, where the mass of the scalar field is assumed to
depend on the local matter density. In such a case fifth-force effects can be strongly 
suppressed on Solar System scales, thus avoiding EP bounds.
Another possibility has been explored in \cite{JMAndre2} where the
authors consider a dilatonic field to be differently coupled to various matter species 
such that the system can naturally evolve toward a late
time attractor solution where General Relativity is recovered.
Non-minimally coupled models can successfully describe the background
expansion of the universe as probed by supernova type Ia luminosity
distance or the position of the Doppler peaks in the Cosmic Microwave
Background anisotropy power spectrum 
(see e.g. \cite{Das,OlivaresAlimiFuzfa}). 
However testing the formation of structure in the universe more than
standard cosmological tests may provide
a key insight on this class of models. 
In fact the scalar coupling contributes
to modifying the clustering properties of matter, implying that an
accurate study of the evolution of density fluctuations both in the
linear and non-linear phase of collapse can identifying unique
signatures of dark sector interactions \cite{PeeblesFarrar}. In the 
context of linear perturbation theory several interacting scalar
field-dark matter models have been studied in the literature 
(see e.g. \cite{DEDMintPert}). In some specific realizations
it was found that the growth of linear density 
perturbations is spoiled by the presence of dangerous
instabilities \cite{Koivisto,Kaplinghat}, as in the case
of ``Mass Varying Neutrino'' (MaVaN) models \cite{Afshordi}.
Recently a number of works have analysed the stability of perturbations in 
more general setups. For instance in \cite{Bean} the authors have studied
models with a background evolution characterized by an adiabatic
regime, and shown that unstable growing modes
of the perturbations exist for couplings much greater than
gravitational strength. On the other hand 
the authors of \cite{Valiviita,Abdalla} have considered the
case of an interacting dark energy component with constant equation of state
and found that for couplings proportional
to the dark matter density the perturbations are unstable.

In this paper we provide a more detailed study of
these instabilities, particularly in relation to the
specificities of the background scalar field evolution. 
The paper is organized as follows: in Section~\ref{DynEq} we introduce
the interacting scalar field-matter model as well as the background and 
perturbation equations; in Section~\ref{PertAnalysis} we present
the results of our analysis; finally in Section~\ref{Conclusions} we 
present our conclusions.

\section{Interacting Scalar Field-Dark Matter Model}\label{DynEq}
Let us consider a scalar field $\phi$ with direct coupling 
to matter particles via a Yukawa term $f(\phi/M_{Pl})\bar{\psi}\psi$,
where $f$ is the coupling function and $\psi$ 
is a Dirac spinor representing the matter field ($M_{Pl}=1/\sqrt{8\pi G}$ 
is the reduced Planck mass with G being the Newton constant).
The effect of the scalar-dependent coupling is to induce a 
time-varying mass of the matter particles, hence causing a violation
of the EP. As mentioned in the
previous Section, there are several ways to evade 
the tight bounds from EP tests. 
Here we assume that the scalar field only couples to dark matter
particles. Therefore for the purposes of our analysis 
we neglect the baryon contribution and focus only on the 
cosmological evolution of the coupled scalar 
field-dark matter system. 

As in the case of the Chameleon cosmology \cite{Brax}, we assume the $\phi$-field to
have a self-interaction potential of runaway type in the form
of inverse power-law:
\begin{equation}
V(\phi)=\frac{M^{4+\alpha}}{\phi^\alpha},\label{invpot}
\end{equation}
where $M$ is a mass scale and $\alpha$ is a positive constant.
We consider a coupling function of dilatonic type, 
$f(\phi)=\exp{(\beta\phi/M_{Pl})}$, with $\beta$ a dimensionless
coupling constant. The background evolution of this system
has been studied in detail in \cite{Das}.

\subsection{Background and Linear Perturbation Equations}
Let assume a flat Friedmann-Lemaitre-Robertson-Walker metric 
($ds^2=-dt^2+a(t)^2d\textbf{x}^2$), the evolution of the scalar factor
is given by:
\begin{equation}
H^2\equiv\left(\frac{\dot{a}}{a}\right)^2=\frac{1}{3}\left[\rho_{DM}+\dot{\phi}^2/2+V(\phi)\right],\label{hub}
\end{equation}
where $\rho_{DM}$ is the dark matter density and we have adopted 
Planck units ($M_{Pl}=1$). The total energy
momentum tensor of the system is conserved, 
$T_{\nu;\mu}^{\mu(T)}\equiv
T_{\nu;\mu}^{\mu(DM)}+T_{\nu;\mu}^{\mu(\phi)}=0$.
In contrast the non-minimal coupling implies that 
the energy momentum tensor of each individual component 
is not conserved. In such a case we can consider
\begin{eqnarray}
T_{\nu;\mu}^{\mu(DM)}&=&\beta \phi_{;\nu}T_\gamma^{\gamma (DM)},\label{tmunu1}\\
T_{\nu;\mu}^{\mu(\phi)}&=&-\beta \phi_{;\nu}T_\gamma^{\gamma (DM)},\label{tmunu2}
\end{eqnarray}
from which we obtain:
\begin{eqnarray}
\dot{\rho}_{DM}+3H\rho_{DM}&=&\beta \dot{\phi} \rho_{DM},\label{rhom}\\ 
\ddot{\phi}+3H\dot{\phi}+V_{,\phi}&=&-\beta\rho_{DM}.\label{KG}
\end{eqnarray}

Without loss of generality we can 
rescale the coupling function $f(\phi)$ to its
present value, $f(\phi_0)$. Hence the solution to
Eq.~(\ref{rhom}) is
\begin{equation}
\rho_{DM}=\frac{\rho_{DM}^{(0)}}{a^3}e^{\beta(\phi-\phi_0)},
\end{equation}
where $\rho_{DM}^{(0)}$ is the present matter density. We may notice
that as consequence of the scalar interaction 
the dark matter density deviates from the standard scaling $a^{-3}$. 
Furthermore for coupling values $\beta>0$, the system of 
Eqs.~(\ref{rhom})-(\ref{KG}) describes an energy transfer 
from the $\phi$-field to dark matter. In such a case
the scalar field evolves in an effective potential
\begin{equation}
V_{\rm eff}(\phi)=V(\phi)+\frac{\rho_{DM}^{(0)}}{a^3}e^{\beta(\phi-\phi_0)},
\end{equation}
which is characterized by the presence of a minimum. 
\begin{figure}[t]
\includegraphics[scale=0.45]{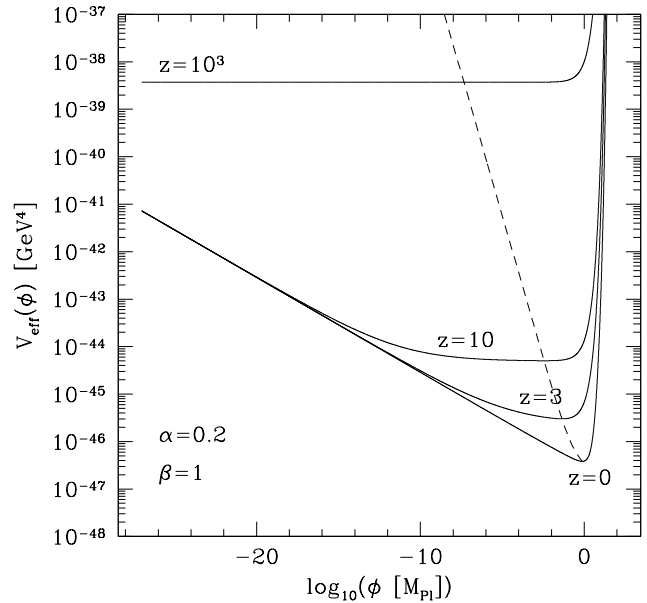}
\caption{Scalar field effective potential at $z=10^3,10,3$
  and $0$ (solid lines) for $\alpha=0.2$ and $\beta=1$. The amplitude of the
  scalar potential $M$ is set such that today $\Omega_{DM}=0.24$ ($\Omega_\phi=1-\Omega_{DM}$).
  The dashed line corresponds to the position of the minimum of the
  effective potential at different epochs.}
\label{fig1}
\end{figure}

In figure~\ref{fig1} we plot the effective potential for $\beta=1$
and $\alpha=0.2$ at redshift $z=1000,10,3$ and $0$ respectively. 
For this choice of the model parameters we have $\phi_0\approx 0.7605$ 
as obtained by integrating numerically the system of 
Eqs.~(\ref{rhom})-(\ref{hub}). The dashed line in Fig.~\ref{fig1}
corresponds to the position of the minimum at different epochs.

In synchronous gauge the linearized equation for the dark matter
density contrast $\delta_{DM}$, velocity gradient $\theta_{DM}$,
and field fluctuation $\delta\phi$ are given by:
\begin{eqnarray}
\dot{\delta}_{DM}&=&-\left(\frac{\theta_{DM}}{a}+\frac{\dot{h}}{2}\right)+\beta
\delta\dot{\phi} \label{deltam},  \\
\dot{\theta}_{DM}&=&-H\theta_{DM}+\beta
\left(\frac{k^2}{a}\delta\phi-\dot{\phi}\theta_{DM}\right) \label{thetam},\\
\delta\ddot{\phi}&+&3 H \delta\dot{\phi}+\left(\frac{k^2}{a^2}+
V_{,\phi\phi}\right) \delta\phi+\frac{1}{2}\dot{h}\dot{\phi}= \nonumber \\
&=&-\beta \rho_{DM}\delta_{DM},
\end{eqnarray}
where $h$ is the metric perturbation given by:
\begin{equation}
\dot{h}=\frac{2 k^2 \eta}{a^2 H}-\frac{8 \pi G}{H}
  \left[\delta\rho_\phi+\rho_{DM}\delta_{DM}\right],
\end{equation}
with $\delta\rho_\phi=\dot{\phi}\delta\dot{\phi}+V_{,\phi}\delta\phi$ and
\begin{equation}
\dot{\eta}=\frac{4\pi G}{k^2}a\left[\rho_{DM}
\theta_{DM}+a k^2 \dot{\phi}\delta\phi\right].\label{eta}
\end{equation}

In Section~\ref{PertAnalysis} we will present the results of the numerical
integration of this system of equations. However for a qualitative
understanding of the conditions which lead to the onset of instabilities
during the growth of the density perturbations, it is useful to
introduce an effective unified fluid description.

\subsection{Effective Unified Fluid Description}\label{effdesc}
The conservation of the total energy momentum tensor allows us to
describe the interacting scalar field-dark matter system
as a single unified fluid. The equation for the background density is
given by
\begin{equation}
\dot{\rho}_{T}=-3H(1+w_T)\rho_T,
\end{equation}
with $\rho_T=\dot{\phi}^2/2+V(\phi)+\rho_{DM}$ and
$w_T=p_T/\rho_T$, where $p_T=\dot{\phi}^2/2+V(\phi)$.
Similarly at linear order the perturbation equations
in synchronous gauge read as:
\begin{eqnarray}
\dot{\delta}_T&=&-3H(c_{sT}^2-w_T)\delta_T+\nonumber\\
&-&(1+w_T)\left\{\left[\frac{k^2}{a^2 H^2}+9(c_{sT}^2-c_{aT}^2)\right]
\frac{a H^2}{k^2} \theta_T+\frac{\dot{h}}{2}\right\},\nonumber \\ \\
\dot{\theta}_T&=&-H(1-3 c_{sT}^2)\theta_T+\frac{c_{sT}^2 k^2}{a(1+w_T)}\delta_T,
\end{eqnarray}
where $c^2_{aT}=\dot{p}_T/\dot{\rho}_T$ is the square of the
adiabatic sound speed of the unified fluid and 
$c^2_{sT}=\delta{p}_T/\delta{\rho}_T$ is the square
of the speed at which pressure perturbations propagate 
in the fluid rest frame. For a barotropic fluid with a constant
equation of state (e.g. matter, radiation) $c^2_s=c^2_a=w$. 
This is not the case for a generic fluid 
(e.g. scalar field), for this reason we may 
expect the effective unified fluid to be non-barotropic, (i.e.
$c^2_{sT}\neq c^2_{aT} \neq w_T$). 
In terms of the scalar field and dark matter variables 
we have
\begin{eqnarray}
c^2_{aT}&=&\frac{3H\dot{\phi}^2+\dot{\phi}[2 V_{,\phi}+\beta
    \rho_{DM}]}{3H\dot{\phi}^2+3H\rho_{DM}},\label{ca}\\
c^2_{sT}&=&\frac{\dot{\phi}\delta\dot{\phi}-V_{,\phi}\delta\phi}{\dot{\phi}\delta\dot{\phi}+V_{,\phi}\delta\phi+\rho_{DM}\delta_{DM}}.\label{cs}
\end{eqnarray}
These relations provide us with a simple way of determining the
properties of the perturbation in the coupled system.
For example in a given background regime 
instabilities of the perturbations may develop if the adiabatic
sound speed acquire sufficiently negative values.

\section{Scalar Field Dynamics and Evolution of Density Perturbations}\label{PertAnalysis}
The non-minimally coupled scalar field
model described in Section~\ref{DynEq} is characterized by
the existence of an attractor solution which is set by 
the minimum of the effective potential.
The minimum is given by
$V^{,\phi}_{\rm eff}=0$, thus along the attractor solution
the following condition is always satisfied:
\begin{equation}
V_{,\phi}=-\beta
\frac{\rho^{(0)}_{DM}}{a^3}e^{\beta(\phi-\phi_0)}.\label{adiacond}
\end{equation}
Evaluating the
derivative of Eq.~(\ref{invpot}) and substituting in Eq.~(\ref{adiacond})
we obtain the time evolution of the field at the minimum:
\begin{equation}
\left(\frac{\phi_0}{\phi_{\rm min}}\right)^{\alpha+1}=\frac{1}{a^3}e^{\beta(\phi_{\rm min}-\phi_0)},\label{fieldadia}
\end{equation}
which depends on both the slope $\alpha$ and the coupling $\beta$. Equation~(\ref{fieldadia}) is a non-linear algebraic equation which can be solved numerically through
standard bisection methods (see dashed line in Fig.~\ref{fig1}).

The field may reach the minimum from two different sets of
initial conditions: $\phi_{ini}<\phi^{ini}_{\rm min}$
(small field) or $\phi_{ini}>\phi^{ini}_{\rm min}$ (large field).
In the former case $\phi$ evolves over the inverse
power-law part of the effective potential, where it minimizes 
the potential by slow-rolling as shown in \cite{Das}. 
In fact one can easily verify that throughout the cosmological
evolution the field mass ($m^2=V_{\rm eff}^{,\phi\phi}$) as well as the
ratio of its kinetic-to-potential energy satisfy the conditions
$m>H$ and $\dot{\phi}^2/2V<1$ respectively. In contrast 
starting from large field values, $\phi$ rolls
towards the minimum along the steep exponential part of $V_{\rm eff}(\phi)$.
Thus it rapidly acquires kinetic energy which subsequently dissipates 
through large high-frequency damped oscillations around the minimum. 

As we shall see next, the growth of linear perturbations in these two regimes 
is significantly different.

\subsection{Adiabatic Regime}\label{adiabaticreg}
As mentioned in Section~\ref{effdesc} we can 
obtain a qualitative insight on the stability of the perturbations 
in the coupled system by considering the effective unified fluid
description. Let us evaluate the adiabatic sound speed
 Eq.~(\ref{ca}) along the adiabatic solution 
Eq.~(\ref{adiacond}); after neglecting 
the term proportional to the kinetic energy of the scalar field we
have
\begin{equation}
c_{aT}^2=-\beta\frac{\dot{\phi}}{3H},\label{cs_adia_cond}
\end{equation}
since $\dot{\phi}>0$ it then follows that $c_{aT}^2<0$, 
implying that adiabatic instabilities may indeed develop. 
However we should remark that during the adiabatic
regime the field is slow-rolling
(i.e. $3H\dot{\phi}\approx 0$), hence 
the term $\dot{\phi}/3H$ can be negligibly small compared 
to $\beta$, such that $c_{aT}^2\approx
0$, hence leading to a stable growth of the perturbations.
In contrast instabilities 
will occur if the coupling assumes extremely large values, 
$\beta\gg 3H/\dot{\phi}$. 
This is consistent with the analysis presented in \cite{Bean}, 
where the authors have suggested that 
during the adiabatic regime perturbations suffer of instabilities
provided that $\beta\gg 1$. Here we want to stress two main
points which were not addressed in that study: 
first of all that the rise of instabilities is suppressed 
by the slow-rolling of the field in the adiabatic regime, and secondly
that exactly because of the slow-roll condition, instabilities can
spoil the growth of dark matter perturbations only for large
unnatural values of the coupling. To give an example let us
assume that for a given model along the adiabatic solution 
the following condition occurs: $\dot{\phi}/3H\sim 10^{-2}$.
In such a case instabilities will develop only if the coupling 
constant $\beta>100$, corresponding to a scalar fifth-force 
which is $2000$ times greater than the gravitational strength.
\footnote{As consequence of the scalar interaction dark matter 
particles experience a gravitational force with effective 
Newtonian constant $G_{\rm eff}=G(1+2\beta^2)$. In contrast baryonic bodies
may not experience such modification due to the non-linear nature of the
scalar interaction \cite{Mota}.} 

Moreover during the adiabatic evolution, 
Eq.~(\ref{cs}) reads as
\begin{equation}
c_{sT}^2=-\frac{1}{1-\frac{1}{\beta}\frac{\delta_{DM}}{\delta\phi}},
\end{equation}
and assuming that the scalar field is nearly homogeneous, 
$\delta\phi\ll\delta_{DM}$ (in Planck units), we have
$c_{sT}^2\approx \beta\delta\phi/\delta_{DM}$,
and for $\beta\approx \mathcal{O}(1)$ this implies $c_{sT}^2\approx 0$.
In other words if the scalar field fluctuations are small
with respect to the dark matter density contrast, 
then the coupled system behaves has
a single adiabatic inhomogeneous fluid ($c_{sT}^2\approx
c_{aT}^2\approx 0$).

These results are supported by 
the numerical study of the system of
Eqs.~(\ref{deltam})-(\ref{eta}), with the scalar field evolution
given by Eq.~(\ref{fieldadia}). We have set the model parameters to the
the following values: $\alpha=0.2$,
$\beta=1$, with $\Omega_{DM}=0.24$, $H_0=70$ ${\rm Km}\,s^{-1}
{\rm Mpc}^{-1}$. 
As shown in \cite{Das} this model has the interesting
feature that the background dynamics can mimic that of a phantom
cosmology corresponding to an uncoupled dark energy model with slightly
constant super-negative equation of state $w_{DE}=-1.1$. 

\begin{figure}[t]
\includegraphics[scale=0.45]{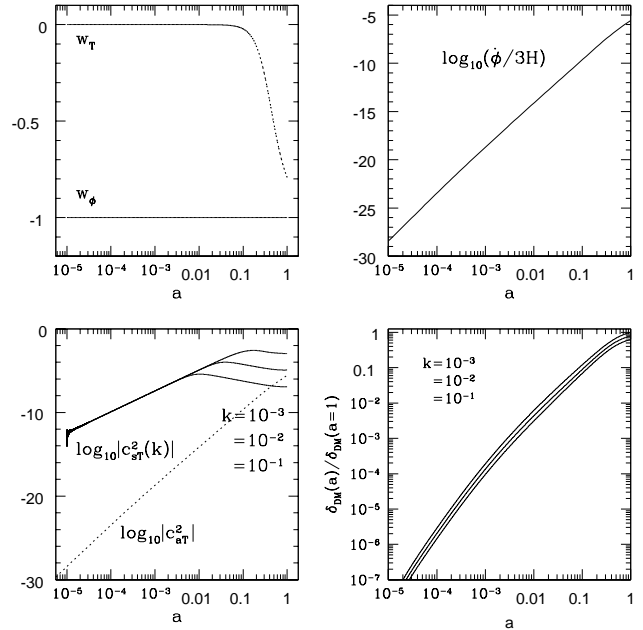}
\caption{Upper left panel: evolution of the scalar field equation of
  state $w_\phi$ and effective unified fluid equation of state $w_T$;
  Right upper panel: evolution of the scalar field velocity with
  respect to the Hubble rate; Lower left panel: redshift evolution of
  the adiabatic sound speed $c_{aT}^2$ and propagation of pressure
perturbations $c_{sT}^2$;
  Right lower panel: Linear growth factor of the dark matter density
  contrast at $k=10^{-3},10^{-2}$ and $0.1$ Mpc$^{-1}$ .}
\label{fig2}
\end{figure}

The results of the numerical integration are shown in Figure~\ref{fig2}. In the upper left panel we
plot the scalar field equation of state $w_\phi$ (solid line) 
and the equation of state for the effective unified fluid $w_T$ (dot
line). As we can see $w_\phi=-1$, which is consistent with the fact that 
$\dot{\phi}/3H$ is negligible, as can be seen from the plot in the right upper panel. 
We can also notice that the unified fluid at early times behaves
as a matter component ($w_T=0$) and 
deviates toward negative values
($-1<w_T<0$) as the $\phi$-field becomes the energetically dominant.
In the lower left panel we plot the absolute value of $c_{aT}^2$, and
$c_{sT}^2(k)$ for three different scales
$k=10^{-3},10^{-2}$ and $0.1$ Mpc$^{-1}$ respectively.
The adiabatic sound speed has negligible negative values and evolves
with a trend that matches that of $\dot{\phi}/3H$, which is consistent with 
Eq.~(\ref{cs_adia_cond}). We can also notice that
the speed of propagation of pressure perturbations 
in the unified fluid remains is $\approx 0$. 
Hence during the adiabatic regime 
the interacting system behaves as a single 
inhomogeneous adiabatic fluid.
In the lower right panel we plot the evolution of the dark matter
density contrast normalized to the present value for
$k=10^{-3},10^{-2}$ and $0.1$ Mpc$^{-1}$ respectively 
(for clarity we have displaced by a constant factor 
the different curves which would otherwise nearly
overlap). As expected these different modes manifest
a standard power law growth and no instabilities are present. 
These results have been obtained for an inverse power-law potential,
nevertheless they can be generalized to other scalar potentials,
the only requirement is the existence of an adiabatic solution during
which the slow-roll condition is satisfied.

\subsection{Non-Adiabatic Regime: Large Field Oscillations}
\begin{figure}[t]
\includegraphics[scale=0.45]{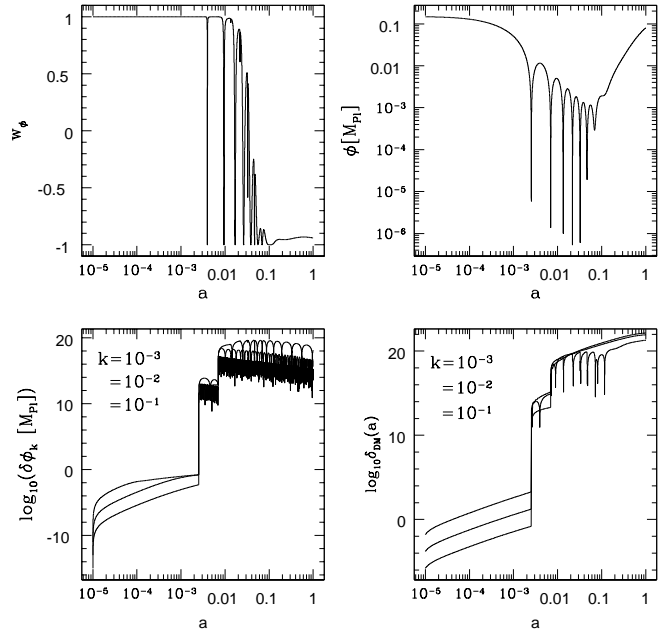}
\caption{Upper left panel: evolution of the scalar field equation of
  state $w_\phi$; Right upper panel: evolution of the scalar field; 
Lower left panel: evolution of the field fluctuations $\delta\phi_{k}$
 at $k=10^{-3},10^{-2}$ and $0.1$ Mpc$^{-1}$ respectively;
  Right lower panel: evolution of dark matter density for $k$-values
  as in the case of $\delta\phi_k$.}
\label{fig3}
\end{figure}

Starting from initially large field values, the system evolves along a  
non-adiabatic solution characterized by rapid dumped field oscillations
around the minimum of the effective potential. We can see this
explicitely the upper left panel of Figure~\ref{fig3}, where we plot
the evolution of the scalar field equation of state for the same
model parameters as in Section~\ref{adiabaticreg} and obtained by 
numerically integrating Eq.~(\ref{KG}) with initial conditions: 
$\phi(a_{ini}=10^{-5})=0.15>\phi^{ini}_{\rm min}$ and $\dot{\phi}_{ini}=0$. 
We can infer the main features of the scalar field evolution from
the behavior of it equation of state shown in the upper left panel of
Figure~\ref{fig3}. As we can see the field initially behaves as a stiff
component with $w_\phi=1$, this is because the field starts rolling on the steep
exponential part of the potential, and consequently its energy is dominated by the
kinetic term. As the field reaches the opposite side of the potential,
it undergoes a series of high-frequency dumped oscillations around the
minimum during which it dissipates most of its kinetic energy. It then sets
on the inverse power-law part of the potential where it evolves along a tracker
solution with $w_\phi\approx -2/(2+\alpha)\approx -0.9$.

The evolution of density perturbations in the case of oscillating scalar
fields has been widely studied in the literature, particularly in
context of inflation \cite{OscillField}. From these
studies it is well known that scalar field fluctuations are unstable
during oscillatory regimes. In \cite{Johson_Kamio} the authors have
presented a simple insightful explanation for the onset of such
instabilities. The idea is to interpret the scalar fluctuation 
$\delta\phi$ as the separation between two particles whose dynamics is
described by two coupled anharmonic oscillators. Then a simple stability 
criterion is given by the relation between the frequency of 
the oscillations $\omega$, and their amplitude $\tilde{\phi}$
\cite{Masso}. Let us suppose that the frequency increases as the
amplitude of the oscillations diminishes, in such a case it has been
shown that the distance between the two particles increases, thus causing
the scalar field fluctuation to be unstable \cite{Johson_Kamio}.
This is indeed what occurs in the interacting scalar field-dark matter
system along the non-adiabatic solution we are considering.
In fact we can see in the right upper panel of Fig.~\ref{fig3} 
that as the field starts oscillating, the frequency of the
oscillations increases as the field amplitude diminishes,
($d\omega/d\tilde{\phi}<0$). We can therefore expect the presence of 
instable modes. This is confirmed by the numerical solutions of $\delta\phi_k$
and $\delta_{DM}$ obtained from the integration 
of Eqs.~(\ref{deltam})-(\ref{eta}). The evolution of the scalar field 
fluctuation $\delta\phi_k$ is shown in the lower left panel of Fig.~\ref{fig3}.  
We may notice the presence of an instability
occurring roughly at the same time of the first oscillation, 
then followed by a second stage of exponential growth at the beginning 
of the second oscillation. From the plot in the lower right panel we can
also see that the same instability is passed to the dark matter
perturbation, which is a direct consequence of the coupling terms
in Eq.~(\ref{deltam}) and Eq.~(\ref{thetam}).
Such unstable modes are similar to those found in
\cite{Valiviita,Abdalla}, in fact by averaging over periods of time larger 
than the characteristic time of the oscillations, the scalar field 
behaves effectively as a dark energy component with a constant equation
of state. 

\section{Conclusions}\label{Conclusions}
We have studied the evolution of linear perturbations 
in the case of an interacting scalar field with runaway potential directly
coupled to dark matter particles. We have specifically analyzed
the stability of perturbations during the adiabatic evolution of the
field, and shown that as consequence of the slow-roll condition
the onsets of instabilities is largely suppressed. 
This can be explained in terms of the adiabatic sound speed 
of the effective unified fluid. In fact during the adiabatic regime,
despite being negative, it assumes negligibly small values and 
as consequence of this the growth of linear density perturbations remains stable. 
On the other hand instabilities may develop in strongly coupled
adiabatic regimes, with a coupling constant much greater 
than gravitational strength. Interestingly during the 
adiabatic evolution of the field the coupled system behaves 
as a single adiabatic inhomogeneous fluid.
We have also shown that large instabilities can spoil the
growth of linear perturbations in the case of non-adiabatic solutions
characterized by large scalar field oscillations. It is well known
that scalar field fluctuations are unstable during oscillatory
regimes, in such a case the scalar coupling amplifies and propagates
such instabilities to the perturbations of the dark matter component. 

Our analysis suggest that under minimal natural model assumptions
Chameleon-like cosmologies are not affected by instabilities of the
perturbations and can provide a viable period of structure
formation more than previously believed. 

\section*{Acknowledgments} 

It is a plaisure to thank Jean-Michel Alimi, Manoj Kaplinghat, Tomi Koivisto, 
Elisabetta Majerotto, David Mota, and Mark Trodden 
for valuable comments and discussions.

\end{document}